\documentclass[aps,prl,twocolumn, groupedaddress,showpacs,showkeys, floatfix]{revtex4-1}
\usepackage{graphicx}
\usepackage[dvipsnames]{xcolor}
\usepackage{amsmath,amssymb,latexsym}
\usepackage[english]{babel}
\usepackage{dcolumn}
\usepackage{bm}
\usepackage{float}
\usepackage[normalem]{ulem}
\usepackage{epstopdf}
\usepackage{mathrsfs}
\usepackage{amsfonts}
\usepackage{subfigure}
\usepackage{wasysym}
\usepackage{soul} 
\usepackage{comment}
\usepackage{enumerate}
\usepackage{helvet}

\newcommand{\va}[1]{{\color{black}{{#1}}}}

\newcommand{\el}[1]{{\color{black}{{#1}}}}

\begin{document}

\title{Globule-like conformation and enhanced diffusion of active polymers}

\author{Valentino Bianco$^{\dag,*}$, Emanuele Locatelli$^{\dag,*}$, Paolo Malgaretti$^{\ddag, \S,}$}

\email{vbianco283@gmail.com, emanuele.locatelli@univie.ac.at,\\ malgaretti@is.mpg.de}

\affiliation{$^\dag$Faculty of Physics, University of Vienna, Vienna, Austria\\
$^\ddag$Max Planck Institute for Intelligent Systems, 
   Heisenbergstr.\ 3, 70569 Stuttgart,   Germany\\
   $^\S$Institute for Theoretical Physics IV,
   University of Stuttgart, Pfaffenwaldring 57, 70569 Stuttgart, Germany}

\date{\today}

\begin{abstract}
We study the dynamics and conformation of polymers composed by active monomers. By means of Brownian dynamics simulations we show that when the direction of the self-propulsion of each monomer is aligned with the backbone, the polymer undergoes a coil-to-globule-like transition, highlighted by a \el{marked} change of the scaling exponent of the gyration radius. Concurrently, the diffusion coefficient of the center of mass of the polymer becomes \el{essentially} independent of the polymer size for sufficiently long polymers or large magnitudes of the self-propulsion. These effects are reduced when the self-propulsion of the monomers is not bound to be tangent to the backbone of the polymer. 
\va{Our results, rationalized by a minimal stochastic model, open new routes for activity-controlled polymer and, possibly, for a new generation of polymer-based drug carriers}.
\end{abstract}

\maketitle



\begin{figure}
\centering
\includegraphics[scale=0.47]{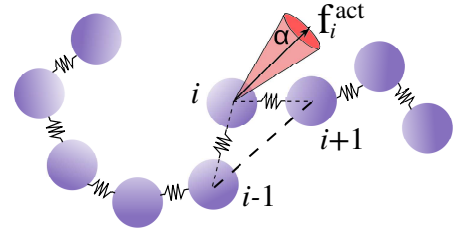}
\caption{Schematic sketch of the active polymer regarded as beads connected by linear springs. For each bead $i$ the active force ${\bf f}_i^{\rm act}$ diffuses in a cone around the vector ${\bf r}_{i+1,i-1}$. The amplitude of the cone is $2\alpha$.}
\vspace{-0.4cm}
\label{fig:model}
\end{figure}

\el{Diverse biological systems feature} chemical reactions and energy conversion \el{occurring} on the backbone of polymers, \el{often involving active components}. For example, DNA is duplicated by DNA-polymerase actively displacing on it~\cite{Albers}; ribosomes synthesize proteins by actively sliding along RNA strands~\cite{Albers,Zia2011}. \el{Syntetic realizations of such processes hint at intriguing applications for micro-devices and nano-medicine~\cite{wang2012nano, dey2015micromotors, medina2015, simmchen2016}. The current state-of-the-art synthesis techniques are already able to mimic biological active filaments with linear chains, composed by active colloidal particles \cite{Dreyfus2005a, Hill2014, Biswas2017, Nishiguchi2018}, often named \textit{active polymers}. From a theoretical perspective,} while many works on the topic have focused on the collective dynamics of \el{active polar gels~\cite{Marchetti2013}, actin filaments~\cite{bathe2008,schaller2010}} and microtubules~\cite{ndlec1997}, recently single-polymer dynamics has received more attention in diverse scenarios spanning from polymers embedded in a bath of active particles~\cite{Kaiser2014, Ghosh2014,Harder2014, Shin2015, Vandebroek2015,  Eisenstecken2016, Samanta2016}, \el{flagellated microswimmers}~\cite{Elgeti2015} and polymers composed by active monomers~\cite{Chelakkot2014, Kaiser2015, Isele-Holder2015, Isele-Holder2016, Osmanovic2017, Winkler2017, gonzalez2018active}.
These works have shown that the details of the coupling between the local active stresses and the conformation of the polymer backbone are crucial for determining the overall dynamics of the polymer. \va{This occurs, for example, in biological processes like DNA-RNA duplication/translation and protein synthesis
where the backbone 
is under the action of tangential forces induced by active displacement of enzymes. }  \\\indent 
In this letter we characterize the structure and dynamics of a self-avoiding linear polymer composed by axis-symmetric active spherical monomers connected by linear springs.
In order to highlight the relevance of the orientation of the activity of the monomers with respect to the local conformation of the polymer we perform Brownian dynamics simulations for different couplings between the local orientation of the active monomers and the conformation of the polymer backbone (Fig~\ref{fig:model}.)
Our results show that when the direction of the axis of the active monomers is tangent to the local instantaneous conformation of the chain, as it happens for ribosomes and DNA-RNA polymerase or for Janus self-propelled \el{necklaces}
~\cite{Biswas2017}, the activity of the monomers reduces the gyration radius of the polymer that enters in a globular-like state.
At the same time, the activity of the monomers promotes the effective diffusion of the polymer inducing an enhanced diffusion coefficient that eventually becomes \el{essentially} independent of the polymer length. 
These effects are due to the tangential action of the active monomers and disappear when the axis of the active monomers is uncorrelated from the conformation of the polymer. In this latter case it has been shown that the activity of the monomers acts as an ``\el{effective higher temperature}''~\cite{Kaiser2015}. 
In order to rationalize our results, we set up a minimal stochastic model that, supported by numerical data, quantitatively captures the dependence of the diffusion coefficient on the controlling parameters. 
\begin{figure*}
\centering
\vspace{-0.7 cm}
\includegraphics[scale=0.24]{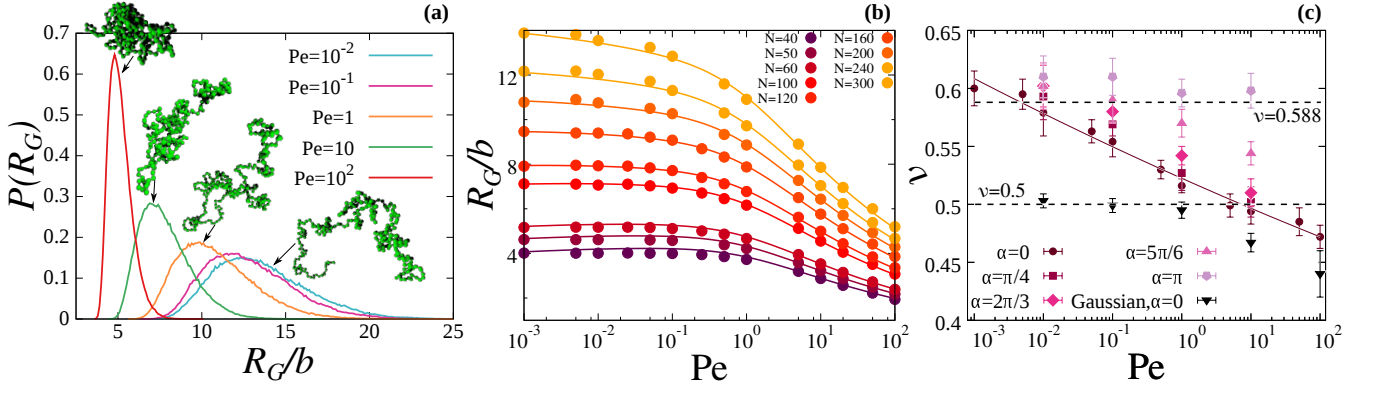}
\vspace{-0.1 cm}
\caption{(a) Probability distribution $P(R_G)$ of the radius of gyration for a polymer $N=300$, for different values of $\rm Pe$. Snapshots of representative polymer conformations at different $\rm Pe$ are shown above the curves.  
(b) $R_G$ as function of $\rm Pe$ for different polymer sizes $N$. \el{Solid lines are fit of the numerical data with Eq. (\ref{eq:def-R_G}), with} $a_{R_G}=0.63$, $h_{R_G}=0.03$ and $c_{R_G} = 0.17$.
(c) Exponent $\nu$ of the gyration radius $R_G$, for different values of $\rm Pe$. Continuous line is a power law fit of $\nu$ for the case $\alpha=0$, given by Eq.(\ref{eq:def-nu}). Similar power-law scaling holds also \el{for $0 < \alpha < \pi$} (and for the Gaussian case for $\rm Pe \geq 1$). Dashed lines show the value of $\nu$ for self avoiding and Gaussian polymers for $\rm Pe=0$. 
}
\label{fig:Rg}
\vspace{-0.25cm}
\end{figure*}

We model the polymer as a bead-spring self-avoiding chain of $N$ monomers in three dimensions, suspended in an homogeneous fluid. The activity of each monomer $i$ is accounted by a force ${\bf f}^{\rm act}_{i}$, with constant magnitude $f^{\rm act}$ that can be made dimensionless by introducing the P\'eclet number ${\rm Pe}$ \footnote{Via the Stokes-Einstein relation, $D_0=\frac{k_B T}{\zeta}$ with $\zeta$ the friction coefficient of the monomer and using $\zeta v^\text{act}\equiv f^\text{act}$ the P\'eclet number in Eq.(\ref{eq:def-Pe}) can be reduced to its common form: ${\rm Pe}=\frac{v^{\rm act} b}{D_0}$.}

\begin{equation}
  {\rm Pe} \equiv \frac{f^{\rm act} b}{k_BT}
  \label{eq:def-Pe}
 \end{equation}
where 
$b$ is the monomer diameter, $k_B$ is the Boltzmann constant and $T$ is the absolute temperature.
\el{We constrain the direction of ${\bf f}^{\rm act}_{i}$ inside a cone with aperture 2$\alpha$,} whose main axes is parallel to ${\bf r}_{i+1,i-1} \equiv {\bf r}_{i+1} - {\bf r}_{i-1}$, i.e the vector connecting the first neighbors of monomer $i$ along the polymer backbone (Fig. \ref{fig:model}). Such construction does not apply to the first and last monomers of the chain, which are passive. In particular, when $\alpha=0$, the vector ${\bf f}^{\rm act}_{i}$ --in a continuous description of the polymer-- is bound to be tangent to the polymer  backbone, inducing a strong correlation between the local stresses induced by the activity of the monomers and the local conformation of the chain. In contrast, for $\alpha=\pi$, each force ${\bf f}^{\rm act}_i $ is independent of the local conformation of the polymer. 

Neighboring monomers along the polymer backbone are held together via a harmonic potential $V^{\rm sp}_i(r) \equiv \sum_{j=i-1,i+1} (K^{\rm sp}/2) \left(r_{i,j} - b  \right)^2$, 
where $r_{i,j}$ is the distance between the monomers $i$ and $j$. Non-neighboring monomers that are closer then the monomer size $b$ repel each other through a purely repulsive harmonic potential 
 $V^{mm}_i(r) \equiv \sum_{j \neq i-1,i+1} (K^{\rm sp}/2) \left(r_{i,j} - b  \right)^2$.  
We fix $K^{\rm sp} = 100$ $k_BT/b$ to avoid crossing events \footnote{During the simulations we check the average monomer-monomer distance, to avoid the over-stretching of the polymer. The model works properly in the range of considered values for $\rm Pe$. For $\rm Pe=10^2$, the average monomer-monomer distance increases $\sim 4.8\%$ of the average value obtained for $\rm Pe=0$.}. We perform Brownian dynamics simulations~\footnote{We use the Euler algorithm, with elementary time step of $dt=10^{-3}$. We have tested our results by decreasing the integration time up to $dt=10^{-5}$ without any quantitative change. Statistics are collected, after equilibration, over up to $10^2$ independent simulations, each of which spans over $10^{10}$ time steps.} integrating the following equation of motion
\begin{equation}
{\bf\dot{r}}_i=\beta D_0\left ( -\boldsymbol{\nabla}_i V_i + {\bf f}^{\rm act}_i\right ) + \boldsymbol{\eta}_i ,
\label{eq:eq_brownian}
\end{equation} 
where $V_i(r) \equiv V^{mm}_i(r) + V^{\rm sp}_i(r)$, $\boldsymbol{\eta}_i $ is a random Gaussian noise satisfying the fluctuation-dissipation relation $\langle {\bf\eta}_{l}(t){\bf \eta}_{k}(t')\rangle = 2 D_0 \, \delta_{l, k}\delta(t-t')$, and $\beta \equiv 1/(k_BT)$.  In the following, we neglect hydrodynamic interactions among monomers, i.e. we investigate the ``Rouse'' regime. As previously mentioned, for $\alpha=0$ the vectors ${\bf f}^{\rm act}_i $ are tangent to the backbone of the polymer, whereas for $\alpha>0$  ${\bf f}^{\rm act}_i$ perform a diffusive motion within a cone of aperture $2\alpha$ according to the equation
\begin{equation}
\dot{\bf f}^{\rm act}_i = \boldsymbol{\eta}_r \times {\bf f}^{\rm act}_i,
\end{equation}
where $D_r$ is the rotational diffusion coefficient satisfying the relation $D_0/D_r = 4/[3 (b/2)^2]$ and the random unit vector $\boldsymbol{\eta}_r $ obeys to the relation $\langle {\bf\eta}_{r_l}(t){\bf \eta}_{r_k}(t')\rangle = 2D_r \, \delta_{l, k}\delta(t-t')$. \el{At} each time step the cone axis is updated and if $ {\bf f}^{\rm act}_i$ exits the cone, it gets bounced back by the exceeding angle. 

First, we consider the effect of the activity on the global \el{conformation} of the chain. In this regard, we compute the radius of gyration $R_G$ as function of the P{\'e}clet number and the polymer size $N$. Interestingly, for $\alpha =0$, we observe a dramatic decrease of the average value of $R_G$; \el{at the same time, the distribution of $R_G$ becomes more peaked} (Fig.~\ref{fig:Rg}a), indicating that the chain gets trapped in a crumpled, collapsed state. 
This behavior, reminiscent of a coil-to-globule transition, is shown in Fig.(\ref{fig:Rg})b,c. $R_G$ can be fitted via a relatively simple function
\begin{equation}
 R_G=b\dfrac{a_{R_G}+h_{R_G}\ln({\rm Pe})}{\left({\rm Pe} +1\right)^{c_{R_G}}}N^{\nu(\rm Pe)}
 \label{eq:def-R_G}
\end{equation}
where $a_{R_G}$, $h_{R_G}$ and $c_{R_G}$ are parameters that are independent of $N$ and $\rm Pe$. 
Interestingly, the good agreement between the prediction of Eq.(\ref{eq:def-R_G}) and the numerical data (see Fig.~\ref{fig:Rg}b) shows that, even for $\text{Pe}\neq 0$, $R_G$ retains a power dependence on $N$, i.e. $R_G \sim N^{\nu(\text{Pe})}$ with 
\begin{equation}
 \nu({\rm Pe}) = 0.52{\rm Pe}^{-0.022} 
 \label{eq:def-nu}
\end{equation}
i.e. $\nu$ diminishes upon increasing $\rm Pe$ (the relation holds for ${\rm Pe} \geqslant10^{-3}$) 
(Figs. \ref{fig:Rg}c and \ref{fig:Rg_alpha}). 
Remarkably, we observe that a similar phenomenology holds in the case of a Gaussian polymer (black triangles in Fig. \ref{fig:Rg}b). This hints that the activity-induced collapse is not strictly related to self-avoidance.
The reduction of $R_G$ upon increasing the activity is \el{surprising}, since the activity has often been suggested to affect the dynamics as an effective \textit{warmer} temperature~\cite{Vandebroek2015,Kaiser2015}. In contrast, our results show the opposite behavior, as the activity leads the polymer towards a globular state, which typically happens upon \textit{cooling} self-attractive polymers. 

Furthermore, as visible in the supplementary videos, the tangent ($\alpha=0$) activity leads the polymer to follow the trail of the first monomers in a sort of ``slithering" motion.
The polymer moves making large, smooth curves, which result in a loose bundle, reminiscent of a common yarn ball. This phenomenon is emphasized by the distribution of the bending angles formed by three consecutive monomers. As shown in Fig.~\ref{fig:p_theta}, upon increasing the value of ${\rm Pe}$ the probability of smaller bending angles increases implying that the polymer is locally more straight. At the same time, larger values of $\rm Pe$ induce more spherical conformations of the polymer (Fig. \ref{fig:asphericity}).

\begin{figure}
\includegraphics[scale=0.3]{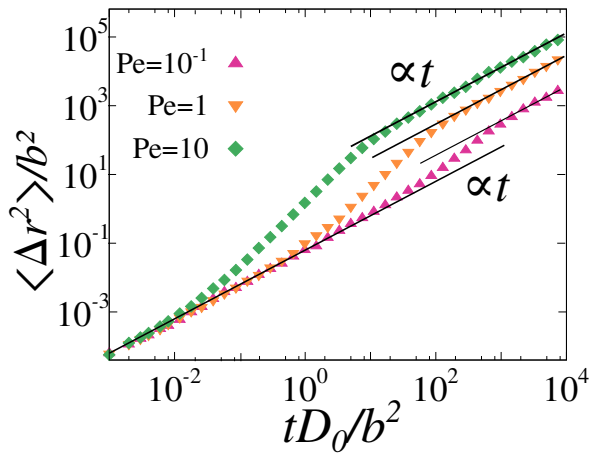}
\vspace{-0.1 cm}
\caption{Mean square displacement $\langle \Delta r^2(t) \rangle$ of the center of mass of the polymer as function of time, for different values of ${\rm Pe}$ and  $N=100$. Black lines are guides for the eye. 
}
\label{fig:msd}
\vspace{-0.15 cm}
\end{figure}

Decoupling the direction of the activity from the conformation of the backbone -- i.e. considering values $\alpha>0$ -- mitigates the collapse of the chain, as captured by the reduction of the scaling exponent of $R_G$ shown in Fig. \ref{fig:Rg}b and Fig.\ref{fig:Rg_alpha}. The decreasing trend in $\nu$ as function of $\rm Pe$, previously discussed for $\alpha=0$, still holds for $0 < \alpha < \pi$, although with a milder slope as $\alpha$ increases. In particular, for $\alpha = \pi$ we recover the conventional scaling exponent $\nu \sim 0.60 \pm 0.02$ for all the values of $\rm Pe$, similarly to what shown in Ref.~\cite{Kaiser2015}. Hence, for $\alpha=\pi$ the activity \el{does not lead to dramatic changes in the polymer conformation}, in contrast to the cases $\alpha\neq\pi$.

Next, we consider the effect of the activity on the dynamics, focusing on the mean square displacement of the center of mass of the polymer $ \langle ({\bf r}_{\rm CM}(t)-{\bf r}_{\rm CM}(0))^2 \rangle \equiv \Delta r^2(t) $.  When $\alpha =0$ three different regimes in the mean square displacement can be identified (Fig. \ref{fig:msd}). At very short times, $\tau<\tau_\text{short}$, a passive diffusive regime $\langle \Delta r^2(t) \rangle \propto D_\text{short}\,t$ with $D_\text{short}=D_0/N$ takes place. At intermediate times, $\tau_\text{short}<\tau<\tau_\text{long}$, a transient super-diffusive regime typical of active systems~\cite{CatesRPP2012}, is observed. Last, at long times, $\tau>\tau_\text{long}$, the diffusive regime is recovered, $\langle \Delta r^2(t) \rangle \propto D_\text{long}\,t$, characterized by an enhanced diffusion coefficient $D_\text{long} > D_\text{short}$. 
We find that $\tau_{\rm short} \propto {\rm Pe}^{-1.5\pm 0.1}$ and that $\tau_\text{short}$ is independent of the polymer size $N$, pointing out that diffusive-to-superdiffusive transition is due to a local dynamics of the monomers (Fig.~\ref{fig:t_cross}a). In contrast, $\tau_{\rm long} \propto N/{\rm Pe}$ i.e, $\tau_{\rm long}$ depends on the global rearrangement of the chain (Fig~\ref{fig:t_cross}b)~\footnote{See Suppl. Mat. for the exact definition of $\tau_\text{short}$ and $\tau_\text{long}$}.


\begin{figure}
\includegraphics[scale=0.19]{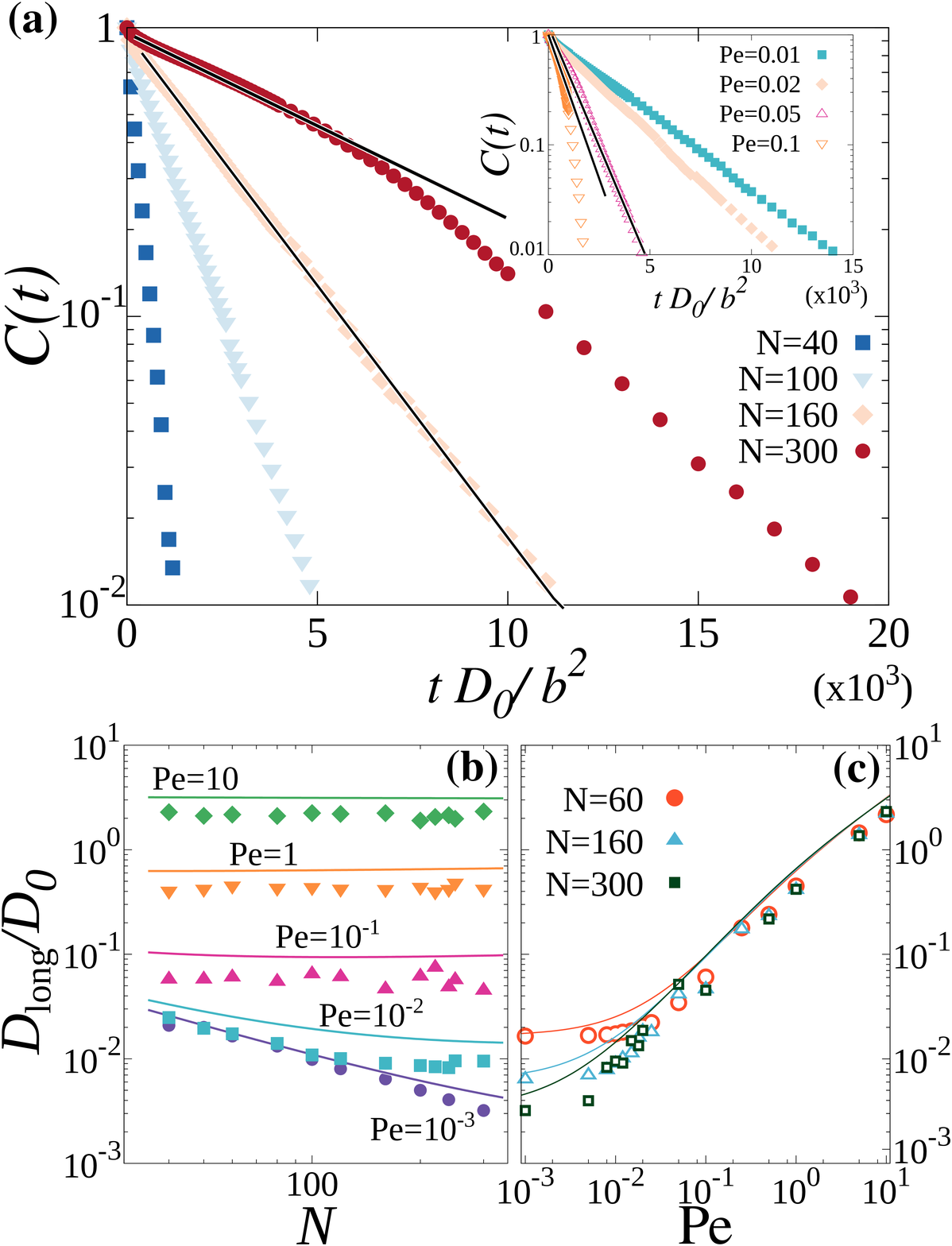}
\vspace{-0.4 cm}
\caption{
(a) Time correlation function $C(t)$ of the end-to-end vector for $\rm Pe=0.02$ and different values of $N$. Black solid lines highlight the exponential decay for lower values of $N$. Inset:  $C(t)$ for $N=160$ and different values of $\rm Pe$.
(b) Translational long-time diffusion coefficient $D_\text{long}$ 
as function of the polymer length $N$, for different values of the P\'eclet numbers $\rm Pe$. 
Data refer to $\alpha=0$. The lines mark the theoretical predictions according to Eq. (\ref{eq:eq_diff}) with $a_{R_E}=1.4$, $h_{R_E}=0.05$ and $c_{R_E}=0.18$. (c) $D_\text{long}$ as function of  $\rm Pe$, for different polymer sizes $N$. The lines mark the theoretical prediction given by Eq. (\ref{eq:eq_diff}). 
}
\label{fig:Ct_D}
\vspace{-0.25 cm}
\end{figure}

Surprisingly, as shown in Fig. \ref{fig:Ct_D}b,c, $D_\text{long}$ becomes independent of the polymer size upon increasing $\rm Pe$. To  rationalize the dependence of $D_\text{long}$ on $N$ and $\rm Pe$ we regard the center of mass of the polymer as a point-like particle under the action of an external force ${\bf F}^{\rm act} \equiv \sum_{i=1}^{N} {\bf f}^{\rm act}_i $, given by the sum of all the contributions stemming from the monomers.
For $\alpha=0$, ${\bf F}^{\rm act}$ is proportional to the end-to-end vector ${\bf R}_E\equiv{\bf r}_{N} - {\bf r}_{1}$.
Accordingly, ${\bf F}^{\rm act}$ can be regarded as a random force  acting on the center of mass with zero average and whose time correlations $C(t)$ are captured by the time correlation of ${\bf R }_E$
\begin{equation}
C(t)\equiv \langle {\bf R}_E(t) \cdot {\bf R}_E(0) \rangle .
\end{equation}
For a passive polymer (${\rm Pe}=0$) the function $C(t)$ decays exponentially \cite{doi_edwards}, with a characteristic relaxation time, $\tau$. In contrast, for ${\rm Pe}\neq0$, Figs. \ref{fig:Ct_D}a shows that the exponential decay holds reasonably well within the range of explored values of $N$ and $\rm Pe$, although for large values of $N$ and $\rm Pe$ a deviation from such a behavior is observed (red circles in Figs. \ref{fig:Ct_D}a and orange triangles in the inset). The behavior of $C(t)$ in such regimes
is consistent with a compressed exponential decay, similarly to what has been observed in
soft-glass and out-of-equilibrium materials~\cite{GabrielJCP2015} where it is due to a
long-range persistent Gaussian noise~\cite{Bouchaud}. This phenomenon goes beyond
the scope of the present paper and will be discussed
in future works.
As shown in the Suppl. Mat. the long-time diffusion coefficient, $D_\text{long}$ can be calculated from the mean square displacement of the center of mass of the polymer that is controlled by the combined action of the thermal noise ($\delta(t)$ correlated in time) of the active force ${\bf F}^{\rm act}$ (correlated in time according to $C(t)$)

\small
\begin{align}
 D_\text{long}\!=\!D_0\!\left\{\frac{D_0 \tau_0}{b^2}\frac{\text{Pe}}{d}\frac{[a_{R_E}+ h_{R_E}\ln(\rm Pe)]^2}{\left(\rm{Pe}+1\right)^{2c_{R_E}}}N^{2\nu({\rm Pe})-1}\!+\!\frac{1}{N}\right\}\! ,
 \label{eq:eq_diff}
\end{align}
\normalsize
where $a_{R_E}$, $h_{R_E}$ and $c_{R_E}$ are fitting parameters that are independent of $N$ and $\rm Pe$ (see Fig.~\ref{fig:scaling_F}b and Eq.~(\ref{eq:def-R_E})), $d$ is the dimensionality of the system and $\nu({\rm Pe})$ is the scaling exponent of $R_E$ (Fig.~\ref{fig:scaling_F}a). 
The first (second) term in Eq.~(\ref{eq:eq_diff}) represents the contribution to the diffusion of the center of mass due to activity (thermal fluctuations). In particular, Eq. (\ref{eq:eq_diff}) shows that for large values of $\rm Pe$ or  $N$ the second term in the brackets can be disregarded and $D_\text{long}$ becomes essentially independent of $N$ since $2\nu({\rm Pe}) -1 \ll 1$~\footnote{Consider that for $\rm Pe=1$ the term $N^{2\nu({\rm Pe}) -1}$ changes of a factor 2 by changing of 3 order of magnitudes the polymer size, from $N=10^2$ to $N=10^5$.}. 
In order to test the reliability of our model we compare the outcome of the numerical simulations and the predictions based on Eq. (\ref{eq:eq_diff}). Interestingly, as shown in Fig.~\ref{fig:Ct_D}b,c our model captures quantitatively the asymptotic growth of $D_\text{long}$ upon increasing $\rm Pe$ whereas it predicts a smoother transition to the passive regime with respect to the numerical results.

\begin{figure}
\centering
\includegraphics[scale=0.35]{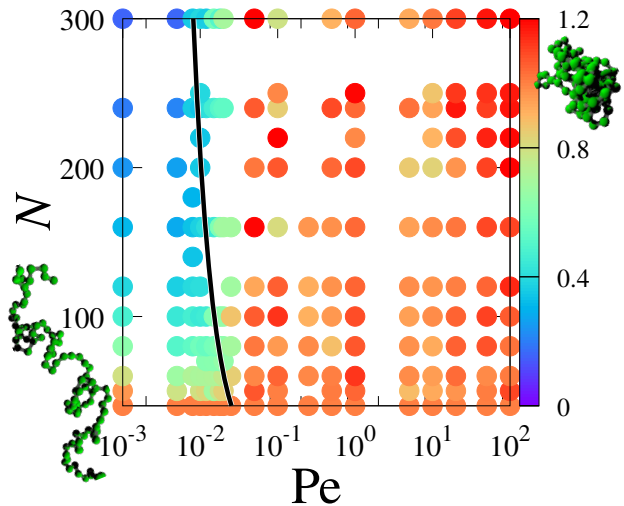}
\vspace{-0.5 cm}
\caption{
Normalized diffusion coefficient ${\tilde D} \equiv D_\text{long}(N,{\rm Pe})/D_\text{long}(40,{\rm Pe})$ (color coded) as function of $N$ and $\rm Pe$. The solid line represents the the locus where the two contributions in Eq.(\ref{eq:eq_diff}) are equals.
}
\vspace{-0.5 cm}
\label{fig:phase_diagram}
\end{figure}

The dynamics of the center of mass of the polymer, \el{as captured by Eq.~(\ref{eq:eq_diff}),} allows us to identify two regions in a $\rm Pe$-$N$ phase diagram, namely a region where $D_\text{long}\propto 1/N$ and a region in which $D_\text{long}$ is almost $N$-independent. Fig.~\ref{fig:phase_diagram} shows the diffusion coefficient $D_\text{long}(N,\text{Pe})$ normalized by the value obtained for $N=40$, for each value of $\rm Pe$. Accordingly, the color code in Fig.~\ref{fig:phase_diagram} indicates the region where $D_\text{long}$ decreases with the polymer size (blue region) and the region where $D_\text{long}$ is almost independent of $N$ (red region). The transition between the two regions, marked by the dashed line in Fig.~\ref{fig:phase_diagram}, is properly captured by our theoretical model.
Moving from the lower to the upper region of the $\text{Pe}$--$N$ phase diagram (Fig.~\ref{fig:phase_diagram}), the radius of gyration and its fluctuations decrease (Fig.\ref{fig:Rg}a), and the continuous coil-to-globule like transition described before can be observed.

For $0 < \alpha < \pi$, a similar phenomenology is observed (Fig \ref{D_alpha}), with a reduced dependence of $D_\text{long}$ on ${\rm Pe}$. In the limiting case $\alpha = \pi$, for which the local activity of the monomer is uncorrelated from the structure of the polymer, a passive-like behavior is recovered, $D_{long} \sim \xi D_0/N$, where $\xi > $ 1 is a prefactor that depends on ${\rm Pe}$ and marks the active nature of the system. In order to address the role of self-avoidance in the aforementioned dynamics we have performed similar numerical simulations for Gaussian polymers  (Fig. \ref{D_gaussian}). We found qualitatively similar results, although with a weaker dependence of $D_\text{long}$ upon $N$ and ${\rm Pe}$. Such a reduced sensitivity is expected since ${\bf F}^{\rm act}\propto \mathbf{R}_E$ that, for a Gaussian polymer, scales with a smaller exponent as compared to a self-avoiding one (Fig.~\ref{fig:Rg}b).

In conclusion, we have studied the dynamics of an active polymer in three dimensions. 
We have shown that both the conformation and the diffusion of the polymer are strongly affected by the activity of the monomers. 
In particular, the effect of the activity 
is strongest when it is bound to be tangent to the backbone of the polymer ($\alpha=0$) and it smoothly reduces upon releasing such a constraint (i.e. increasing $\alpha$). 
Concerning the polymer conformation, we found that when the activity dominates over the thermal motion, the polymer undergoes a coil-to-globule-like transition as captured by the 
\va{decrease of the}  
scaling exponent of the gyration radius (Fig.\ref{fig:Rg}), i.e. \va{ increasing the activity is analogous to \emph{reducing} the temperature for self-attracting polymers.
At the same time, }
the diffusion coefficient of the polymer becomes independent of its size and larger than the corresponding equilibrium value. In this \el{latter} respect the activity acts as a \emph{higher} temperature that enhances the diffusion.
These results might open the route for highly mobile drug delivery carriers made out of active polymers. Indeed, current state-of-the-art techniques may open up the possibility to synthesize active polymers whose active monomers have their axis of motion aligned with the polymer backbone, by means of surface-shell functionalization of single colloids \cite{Liu2010a, VanderMeulen2013,Feng2013}.



\vspace{0.2 cm}

We acknowledge I. Coluzza, C. Dellago, C.N. Likos, L. Rovigatti for helpful discussions. V. B. acknowledges the support from the Austrian Science Fund (FWF), Grant No. M 2150-N36. The computational results presented have been achieved using the Vienna Scientific Cluster (VSC).

\bibliography{biblio.bib}

\cleardoublepage

\onecolumngrid

\appendix

\renewcommand\thefigure{\thesection A.\arabic{figure}} 
\setcounter{figure}{0}

\section{SUPPLEMENTARY MATERIAL}

\subsection{Structural properties} 

\begin{figure}[h!]
\leftskip 5.5 cm
\large
$\rm Pe=10^{-1}$ \hspace{4.5 cm} $\rm Pe=1$\\ \normalsize
\centering
\includegraphics[scale=0.28]{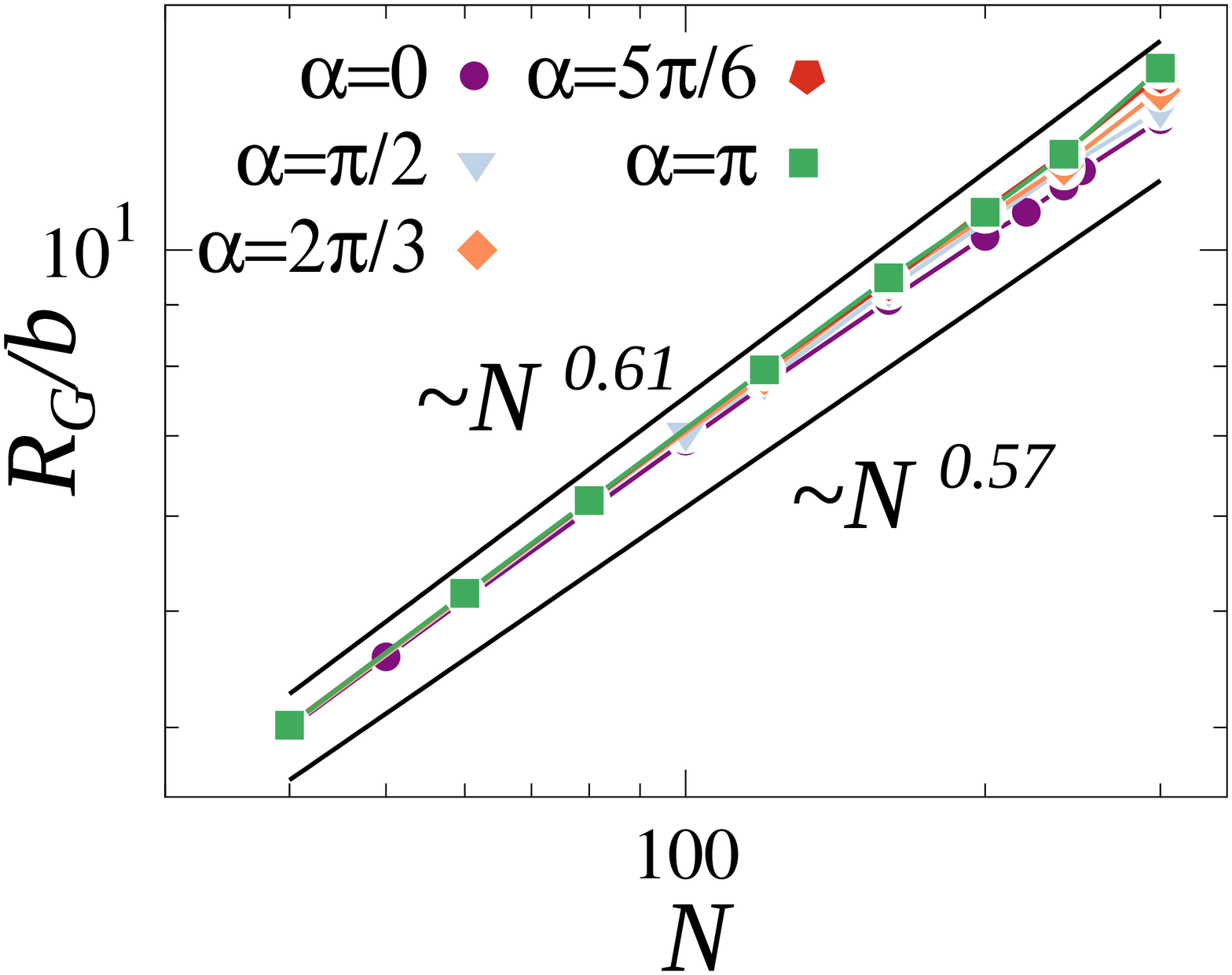}\hspace{0.1 cm}
\includegraphics[scale=0.28]{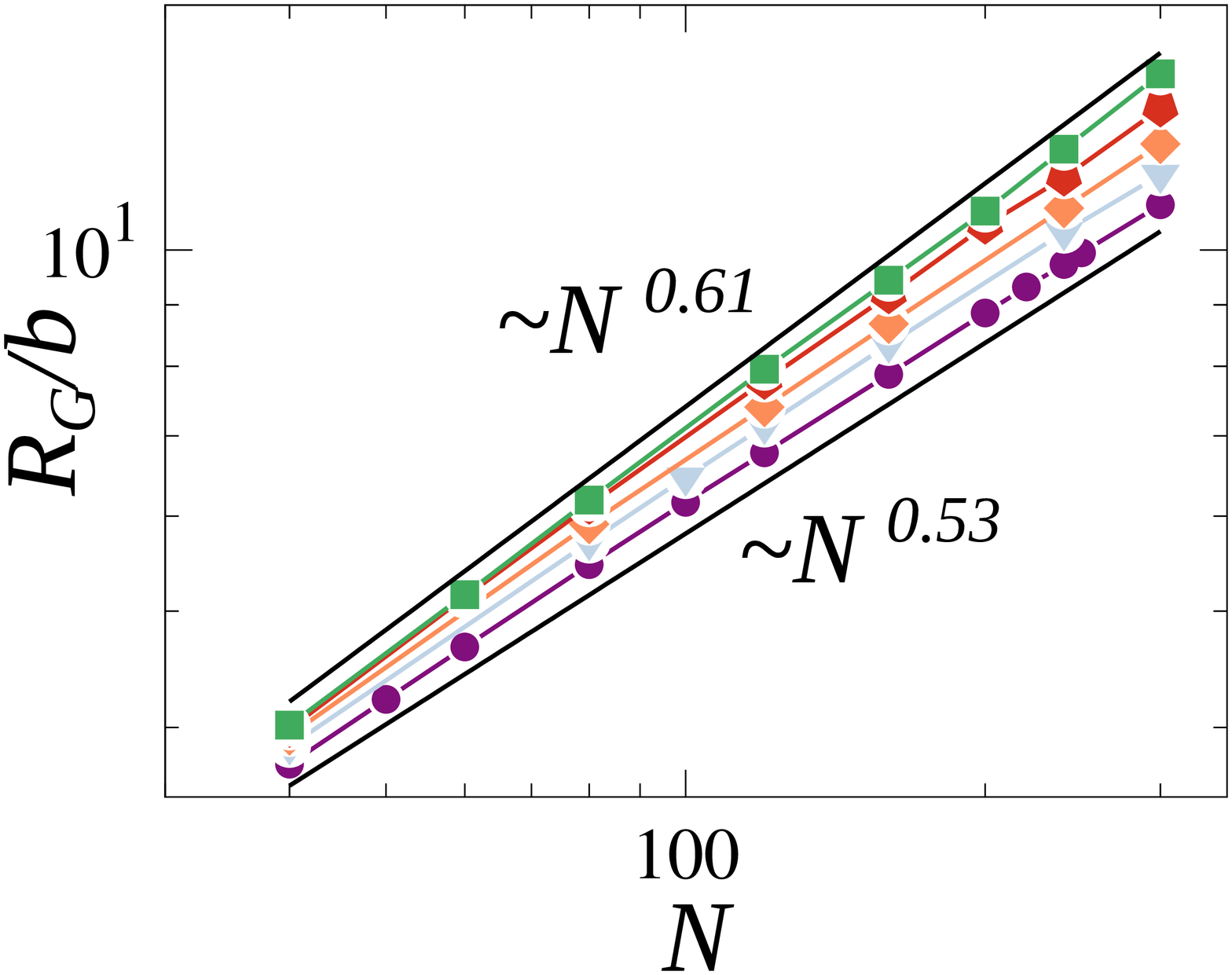}\\
\vspace{0.1 cm}
\leftskip 5.5 cm
\large
$\rm Pe=10$ \hspace{4.5 cm} $\rm Pe=10^2$\\ \normalsize
\centering
\includegraphics[scale=0.28]{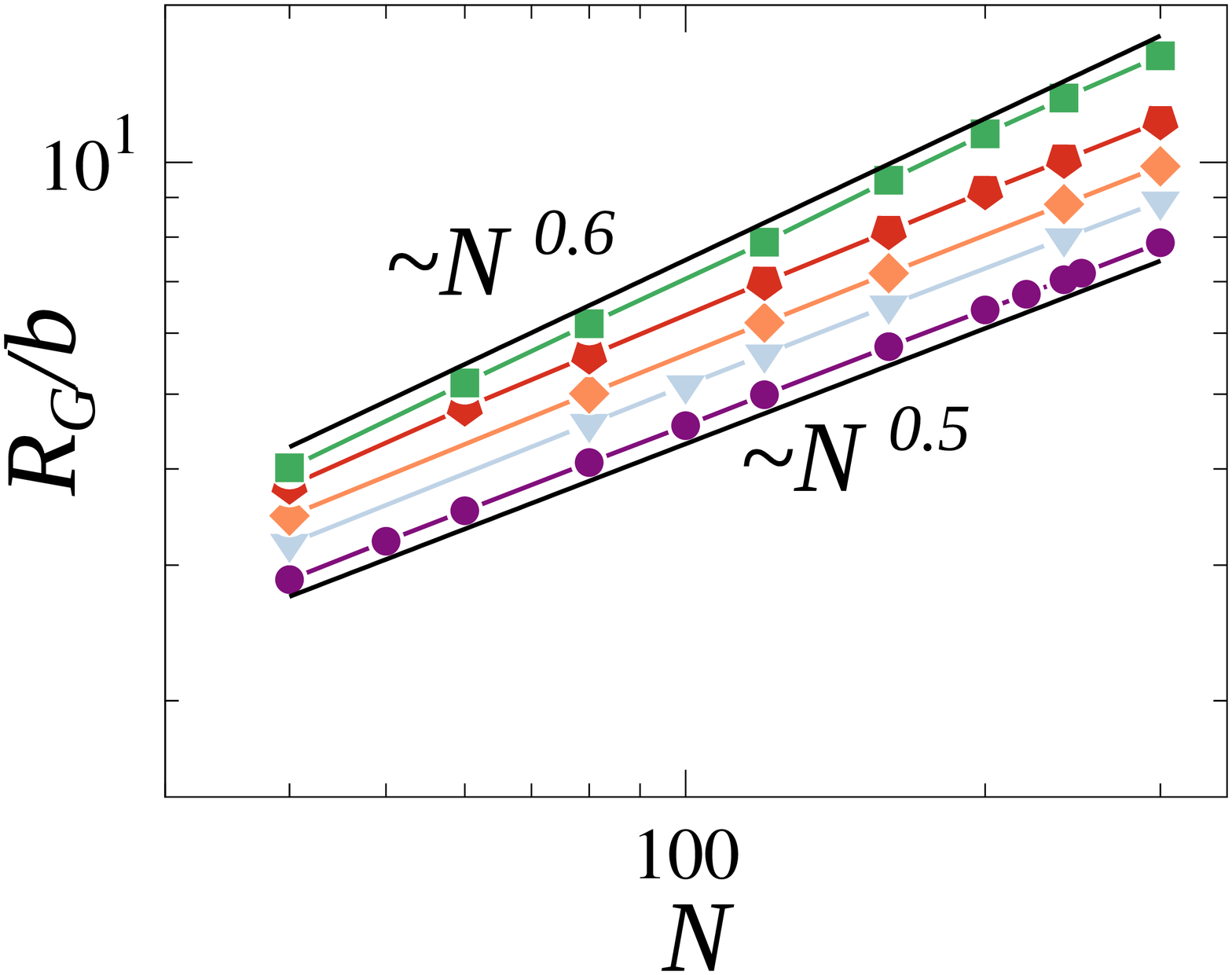}\hspace{0.1 cm}
\includegraphics[scale=0.28]{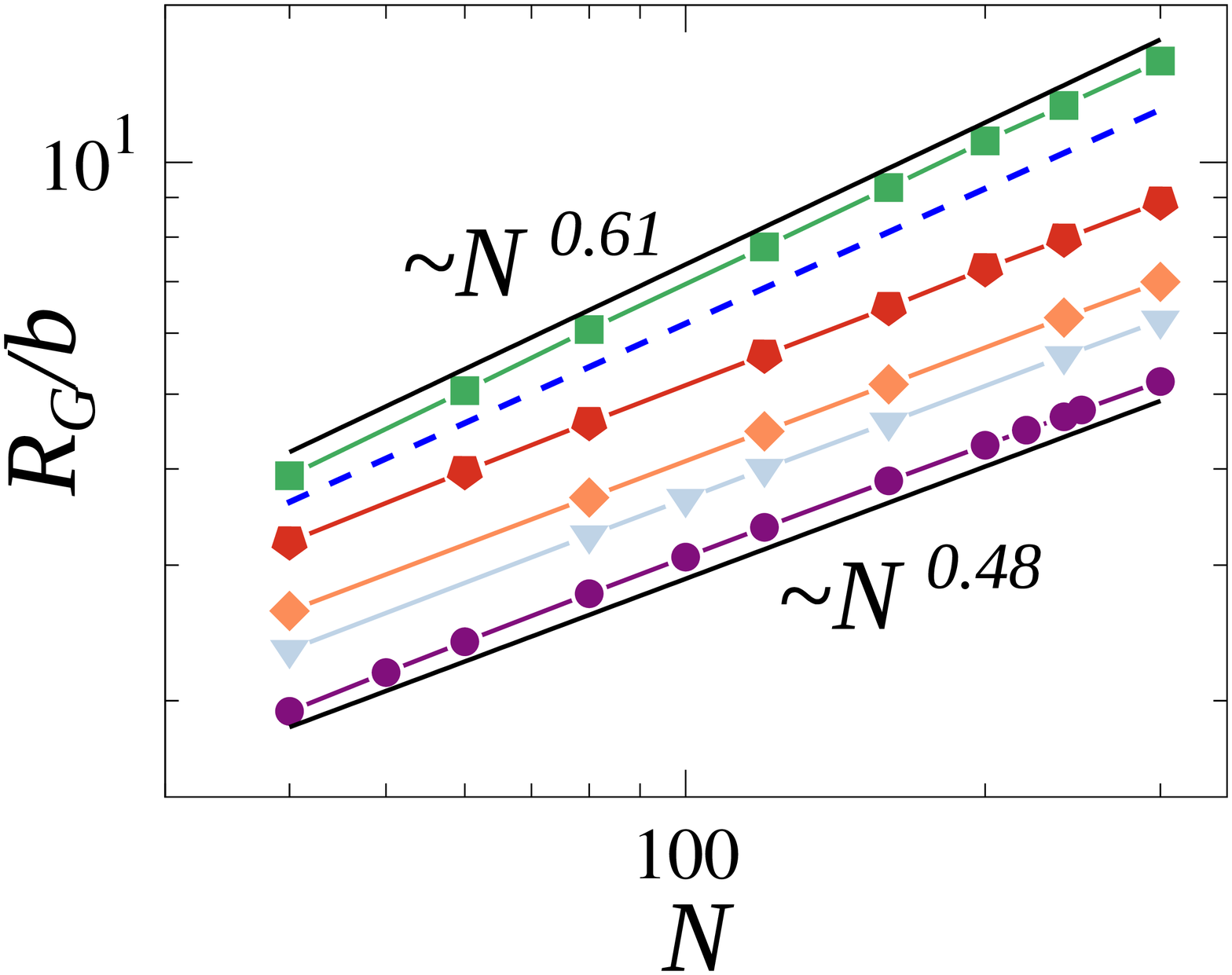}\\
\caption{Radius of gyration $R_G$ as function of $N$, for different values of $\alpha$. Black lines mark the scaling behavior of the lower curves (always corresponding to the case $\alpha=0$) and the upper curves (always corresponding to the case $\alpha=\pi$), and are shifted respectively down and up for sake of visualization. Dotted blue line in the last panel show the scaling behavior of a passive polymer $\sim N^{0.59}$.}
\label{fig:Rg_alpha}
\end{figure}

In Fig.~\ref{fig:Rg_alpha} we report the radius of gyration as function of the length of the polymer, $N$, for different cone aperture $\alpha$ (different symbols) and different P\'eclet numbers ${\rm Pe}$ (different panels). At the lowest ${\rm Pe}$, for $\alpha < \pi$ the shrinking of the chain is barely visible; increasing ${\rm Pe}$ the effect becomes more and more evident, leading to different scaling exponents for chains with different values of $\alpha$. 

 \begin{figure}[h!]
  \includegraphics[scale=0.35]{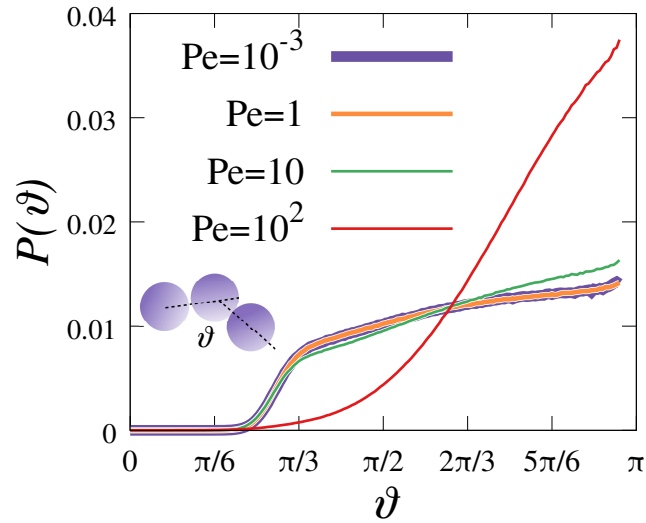}
  \caption{Probability distribution $P(\theta)$ of the angle $\theta$ formed between three contiguous monomers for different values of $\rm Pe$. A scheme, defining the angle $\theta$ is shown in the inset. 
  The globular structure of the polymer for larger values of $\rm Pe$ is emphasized by larger values of $P(\theta)$ for bigger angles.
  Data are calculated for $N=300$ and $\alpha =0$. }
  \label{fig:p_theta}
  \end{figure}
Next, we report the probability distribution $P(\vartheta)$ of the angle $\vartheta$ formed between neighboring monomers. For a passive polymer, this quantity is almost flat for $\vartheta > \frac{2\pi}{3}$ and zero otherwise. This is a known effect of excluded volume interactions, which greatly penalize configurations with partial overlap and induces an "effective" bending rigidity. In contrast, for larger values of ${\rm Pe}$ larger angles, i.e. straighter local configurations, become predominant. This is a confirmation of the scenario discussed in the main text, as curves and bends in a yarn bundle configuration are characterized mostly by  larger local curvature. 

\begin{figure}
 \includegraphics[scale=0.35]{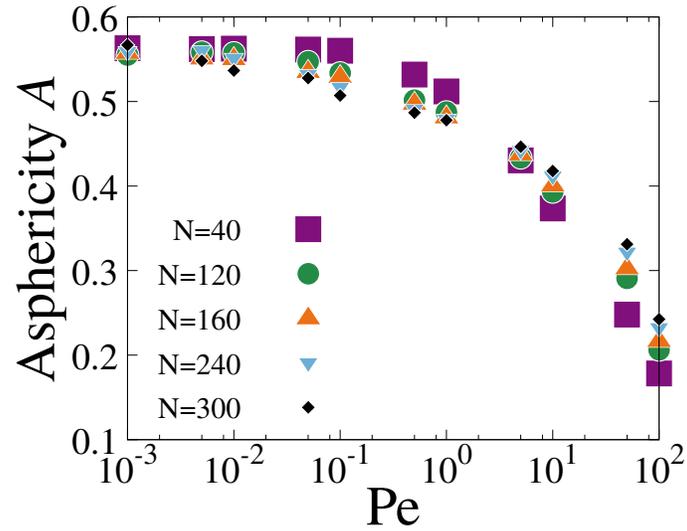}
  \caption{Asphericity $A$ of the polymer as function of $\rm Pe$ for different values of $N$. Data are calculated for $\alpha=0$. }
  \label{fig:asphericity}
\end{figure}

Finally we compute the asphericity $A$ of the polymer, defined as

\begin{equation}
 A = \dfrac{\langle {\rm Tr}^2 - 3M \rangle }{ \langle {\rm Tr}^2 \rangle }
\end{equation}
where $\rm Tr\equiv \lambda_1 + \lambda_2 + \lambda_3$, $M\equiv \lambda_1\lambda_2 + \lambda_1\lambda_3 +  \lambda_2\lambda_3$, and $\lambda_i$ with $i=1,2,3$ are the three eigenvalues of the gyration tensor. The symbol $\langle ... \rangle$ indicates the statistical average. The asphericity ranges from 0 for a spherical conformation, to 1. In Fig. \ref{fig:asphericity} we show $A$ as function of ${\rm Pe}$ for different polymer sizes $N$. We find that the activity affects the geometry of the polymer leading to more spherical conformations for higher values of $\rm Pe$. This effect does not depend on the polymer size $N$.

\newpage
\subsection{Dynamical properties}

\begin{figure}[h!]
\leftskip 4.9 cm
\large
$\rm Pe=10^{-1}$ \hspace{5.5 cm} $\rm Pe=10$\\ \normalsize
\centering
\includegraphics[scale=0.3]{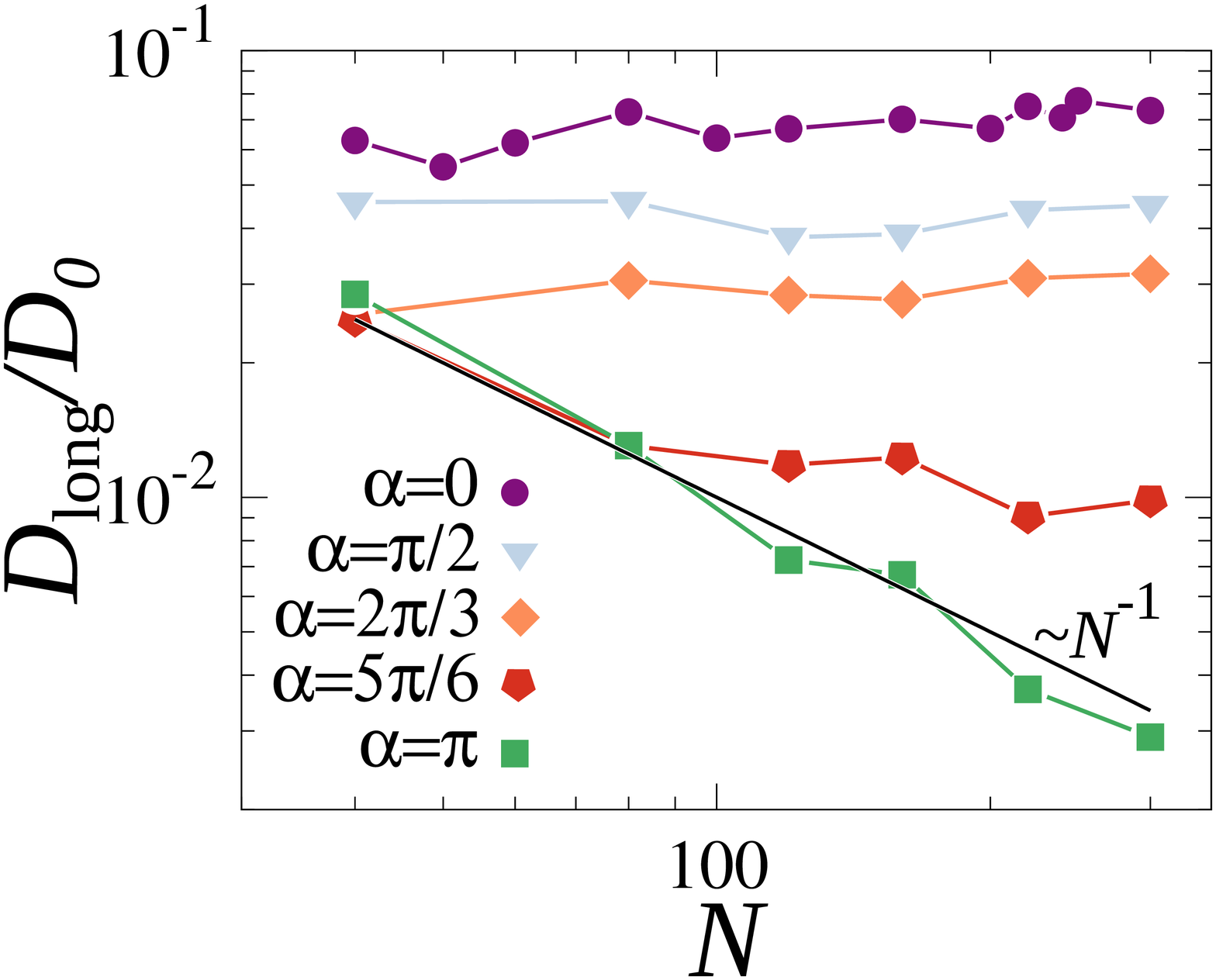}\hspace{0.5 cm}
\includegraphics[scale=0.3]{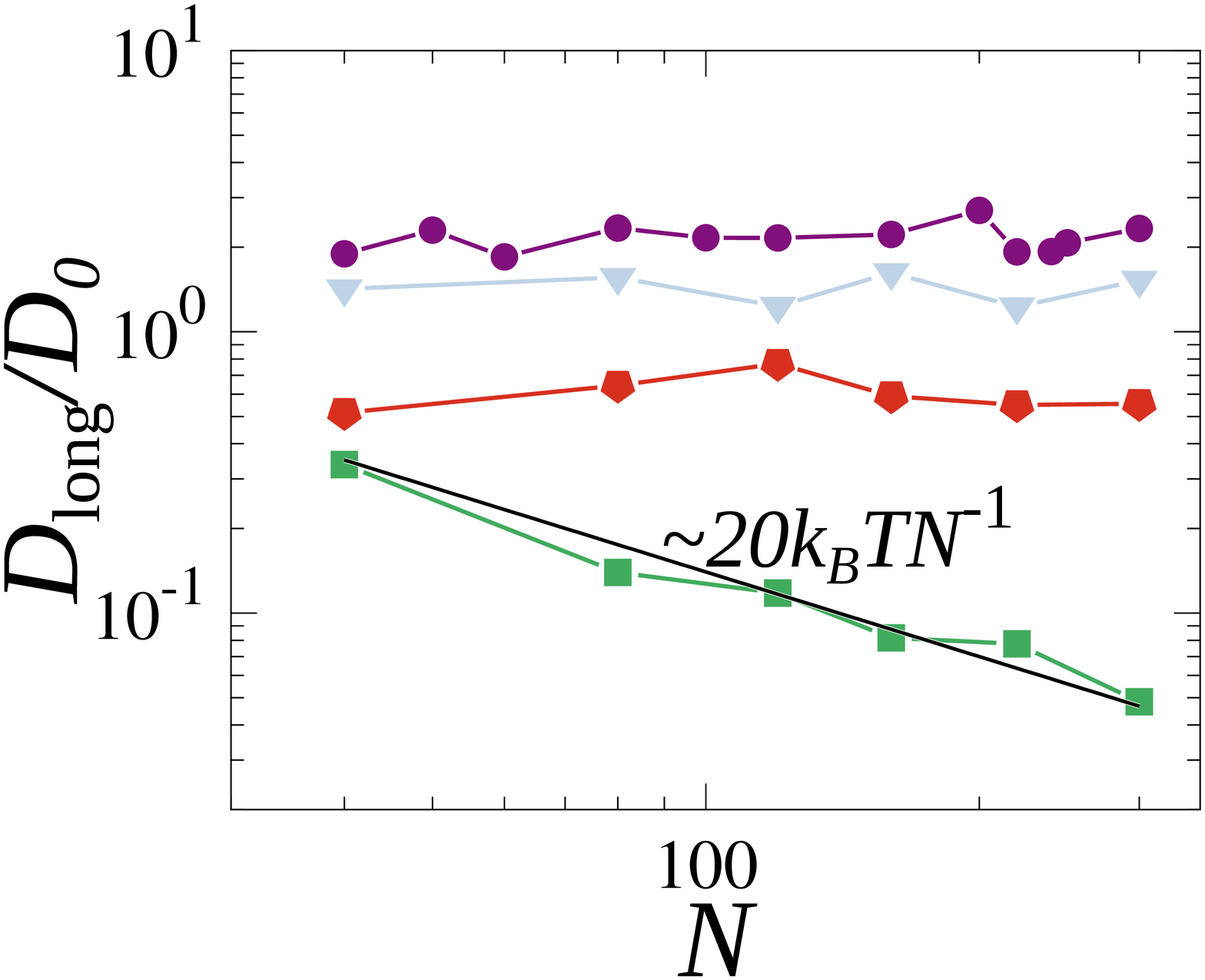}\\
\caption{Diffusion coefficient $D_\text{long}$ as function of $N$, for different values of $\alpha$. $\alpha=0$ corresponds to ${\bf f}^{\rm act}$ tangent to the backbone; \el{$0 < \alpha < \pi $ corresponds to ${\bf f}^{\rm act}$ directed in a cone of amplitude $2\alpha$, whose axis coincides with the vector tangent to the backbone} (Fig. \ref{fig:model}). $\alpha=\pi$ corresponds to ${\bf f}^{\rm act}$ randomly distributed. For $\rm Pe = 10^{-1},1,10$ the black line, overlapping with the green points related to the case $\alpha=\pi$, scales as $k_BT/N$. For $\rm Pe = 10$ the black line scales as $ \sim 20k_BT/N$.}
  \label{D_alpha}
\end{figure}

In Fig.\ref{D_alpha}, we report the long-time diffusion coefficient $D_\text{long}$ of the center of mass of the chain as function of $N$ for two representative values of ${\rm Pe}$ and four different values of the cone aperture $\alpha$ (Fig.~\ref{D_alpha}). We notice that, at fixed ${\rm Pe}$, as long as $\alpha < \pi$, i.e. as long as the activity of the monomers is correlated to the local configuration of the polymer, the qualitative trend remains similar to that observed for $\alpha =0$ but with a reduced magnitude. 
For $\alpha = \pi$, the passive-like behavior is restored although, due to the activity, the diffusion coefficient is larger than the purely passive one. 
This suggests that upon increasing $\alpha$ the net effect of activity on polymer diffusion is hindered. 

\begin{figure}[h!]
\vspace{-0.2 cm}
\includegraphics[scale=0.31]{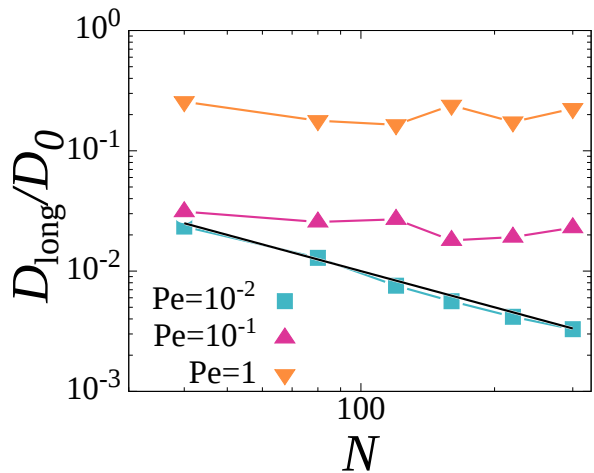}
\vspace{-0.3 cm}
\caption{Long-time diffusion coefficient $D_\text{long}$ for an active Gaussian polymer as function of $N$, for different values of $\rm Pe$. Black line shows the behavior of a passive Gaussian polymer in the bulk.}
\label{D_gaussian}
\end{figure}

In Fig.\ref{D_gaussian}, we report the long-time diffusion coefficient $D_\text{long}$ of the center of mass of a Gaussian chain as function the chain length $N$ for different P\'eclet numbers and for $\alpha = 0$. The behavior is qualitatively similar to that observed for a self-avoiding chain. In particular, for a Gaussian polymer the magnitude of the effect is smaller, as expected since the active force scales with a smaller exponent as compared to the self-avoiding case. 

Finally, we show the scaling behavior of the crossover times $\tau_{\rm short}$ and $\tau_{\rm long}$, which respectively mark the transition from the diffusive to the super-diffusive regime at shorter times, and the transition from the super-diffusive to the diffusive regime at longer times (Fig. \ref{fig:msd}). To compute $\tau_{\rm short}$ and $\tau_{\rm long}$ we first fit the mean square displacement data with a function $\langle \Delta r^2(t) \rangle = D_{\rm short} t$ for shorter times, where $ D_{\rm short}$ represents the diffusion coefficient at shorter times. Then we fit the superdiffusive regime with a function $\langle \Delta r^2(t) \rangle = D_{\rm super} t^\varepsilon$, where $D_{\rm super}$ and $\varepsilon$ are fitting coefficients. In particular, $\varepsilon$ marks the non-linear scaling of the mean square displacement with the time, and we find $\varepsilon=2.0\pm0.2$ for all the data. Finally, at longer times we fit our data with the function  $\langle \Delta r^2(t) \rangle = D t$, where $D_\text{long}$ is the diffusion coefficient shown in Fig. \ref{fig:Ct_D}b. The two  intersection points between the three fitting curves identify $\tau_{\rm short}$ and $\tau_{\rm long}$.

\begin{figure}
 \includegraphics[scale=0.35]{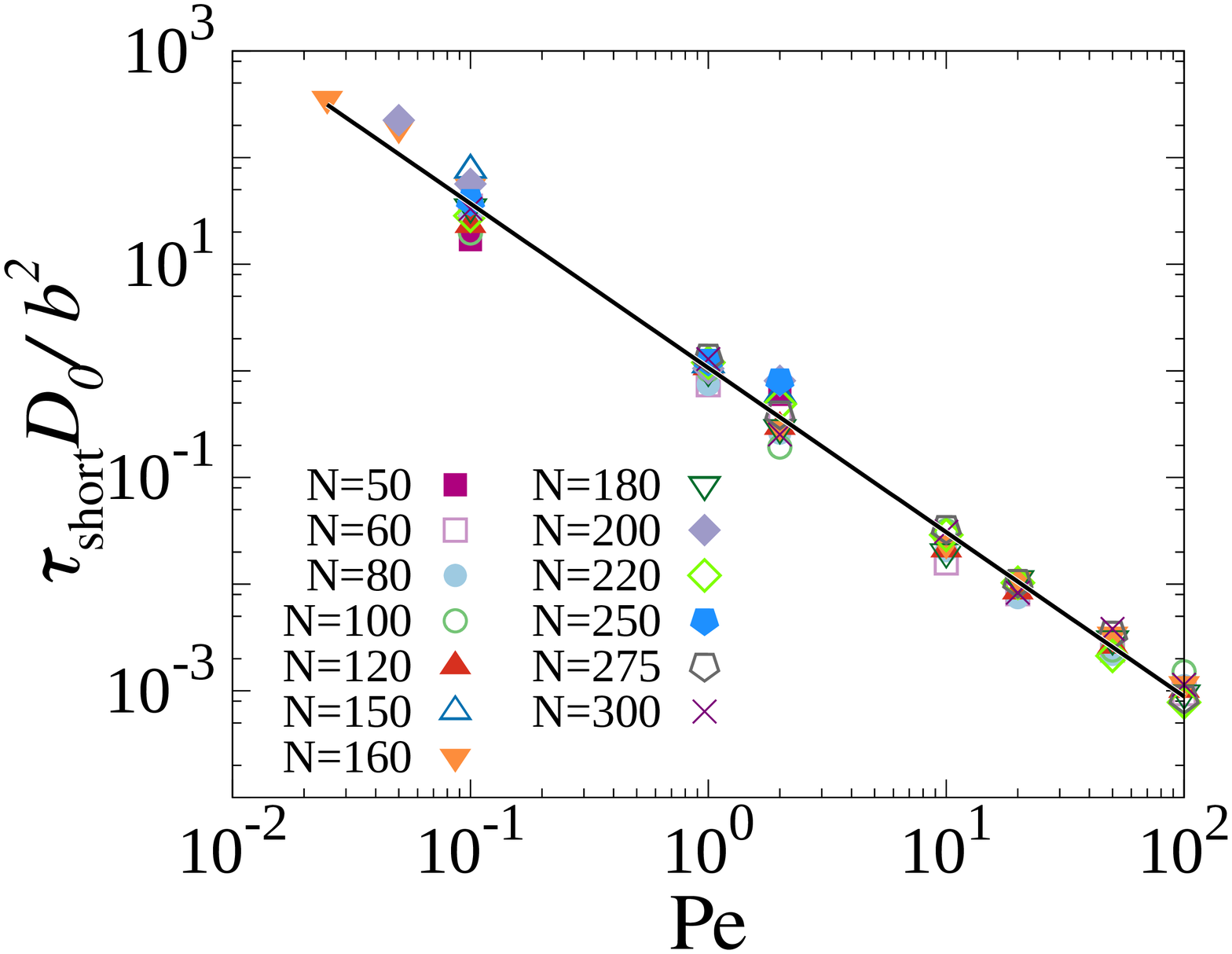}
  \includegraphics[scale=0.35]{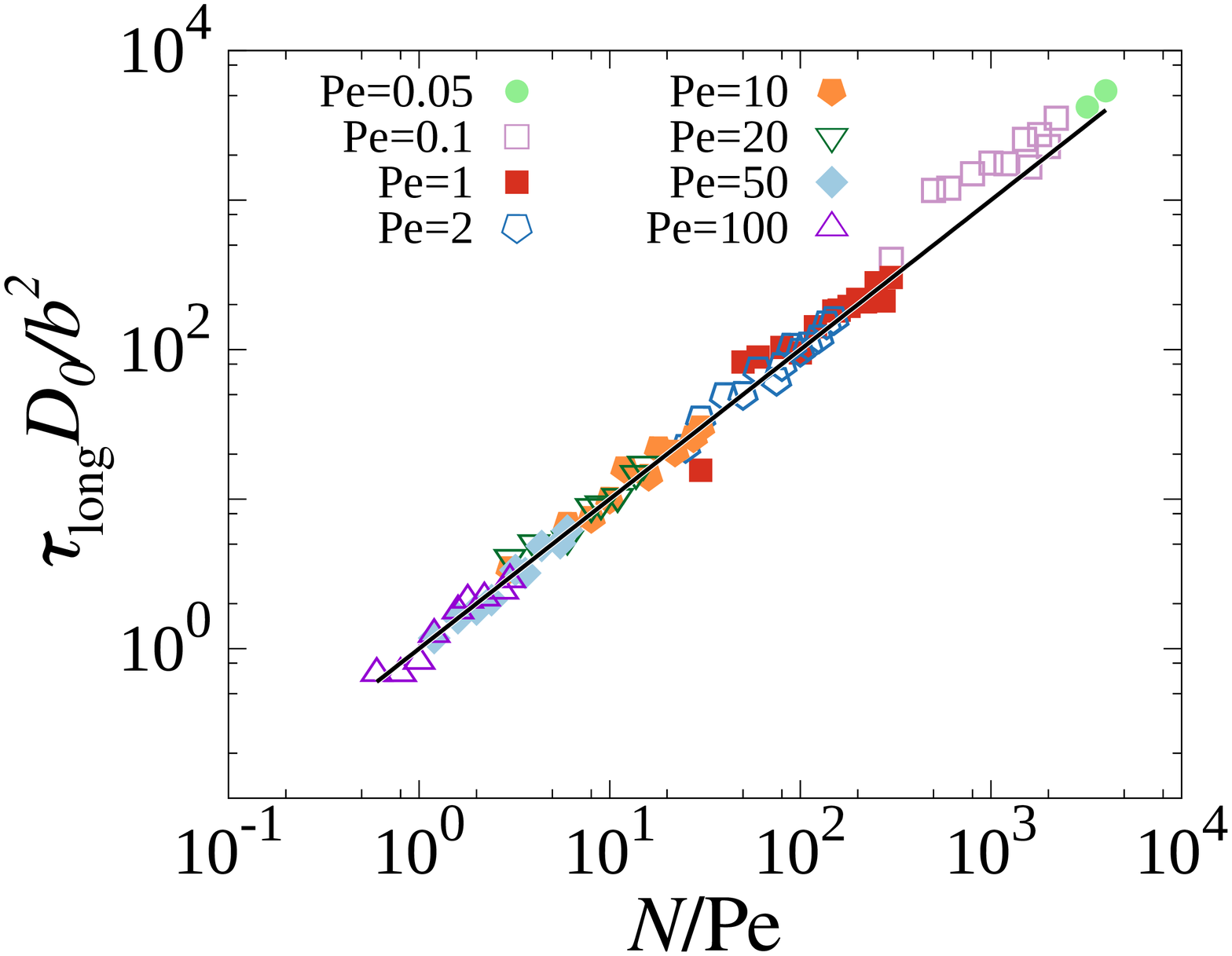}
 \caption{Scaling of the crossing time $\tau_{\rm short}$. The curve shown is $\tau_{\rm short} = {\rm Pe}^{-3/2}$. (b) Scaling of the crossover time $\tau_{\rm long}$. All the data collapse on the curve $\tau_{\rm long}=N/{\rm Pe}$.}
 \label{fig:t_cross} 
\end{figure}

\subsection{Theoretical approach}
\paragraph{Active force on the center of mass.}
In the following we derive the effective diffusion coefficient of the center of mass of a polymer composed by active monomers (Eq.~(5) of the main text). We consider the case where the active force ${\bf f}_i^{\rm act}$ acts along the monomer bonds, i.e. $\alpha=0$. In this case the total active force acting on the center of mass 
\begin{equation}
 \mathbf{F}^{\rm act}=f^\text{act}\sum_{i=2}^{N-1}\dfrac{\mathbf{r}_{i+1}-\mathbf{r}_{i-1}}{|\mathbf{r}_{i+1}-\mathbf{r}_{i-1}|}
 \label{eq:def-F0}
\end{equation}
In a continuum representation of the polymer, \el{the total active force is proportional the integral over the polymer backbone of the unit tangent vector $\hat{t}$,} namely:
\begin{equation}
 \mathbf{F}^{\rm act}=\frac{f^\text{act}}{b}\int\limits_0^L \hat{\mathbf{t}}dl=\frac{f^\text{act}}{b}\mathbf{R}_E
 \label{eq:def-F}
\end{equation}
\begin{figure}
\leftskip 1.3 cm {\bf (a)\hspace{5 cm} (b) \hspace{5.3 cm} (c) } \\
\centering
\includegraphics[scale=0.24]{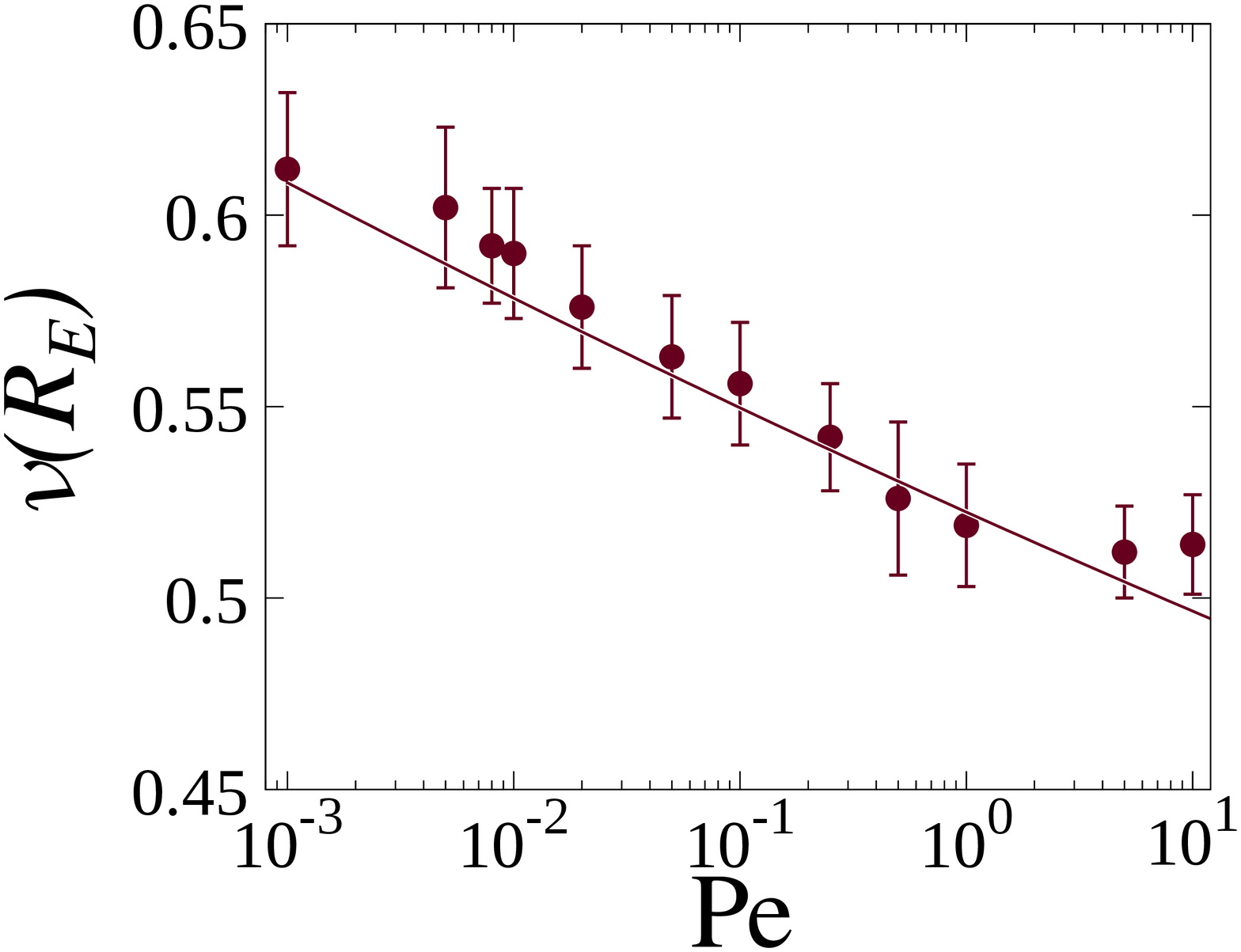} 
 \includegraphics[scale=0.24]{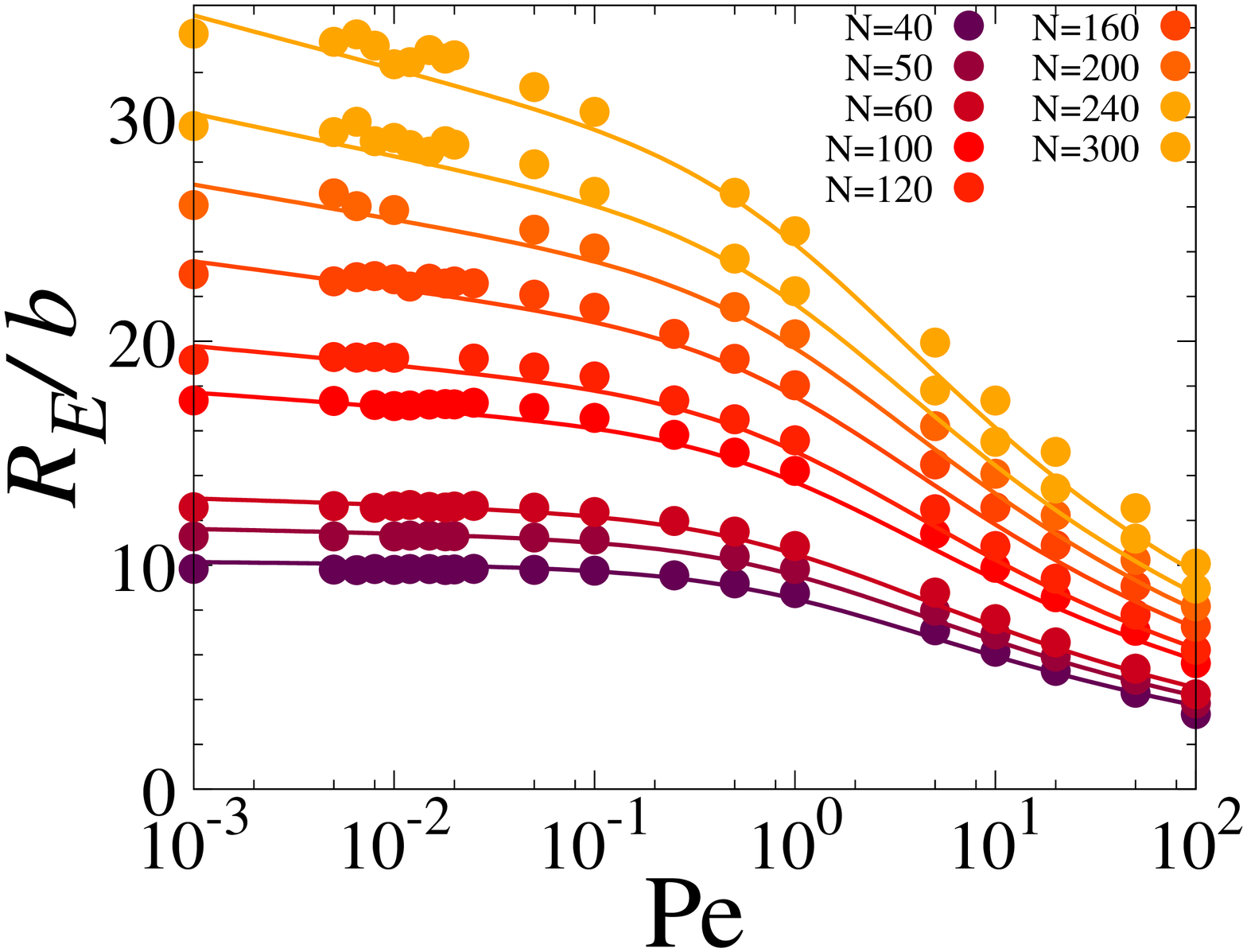} 
 \includegraphics[scale=0.24]{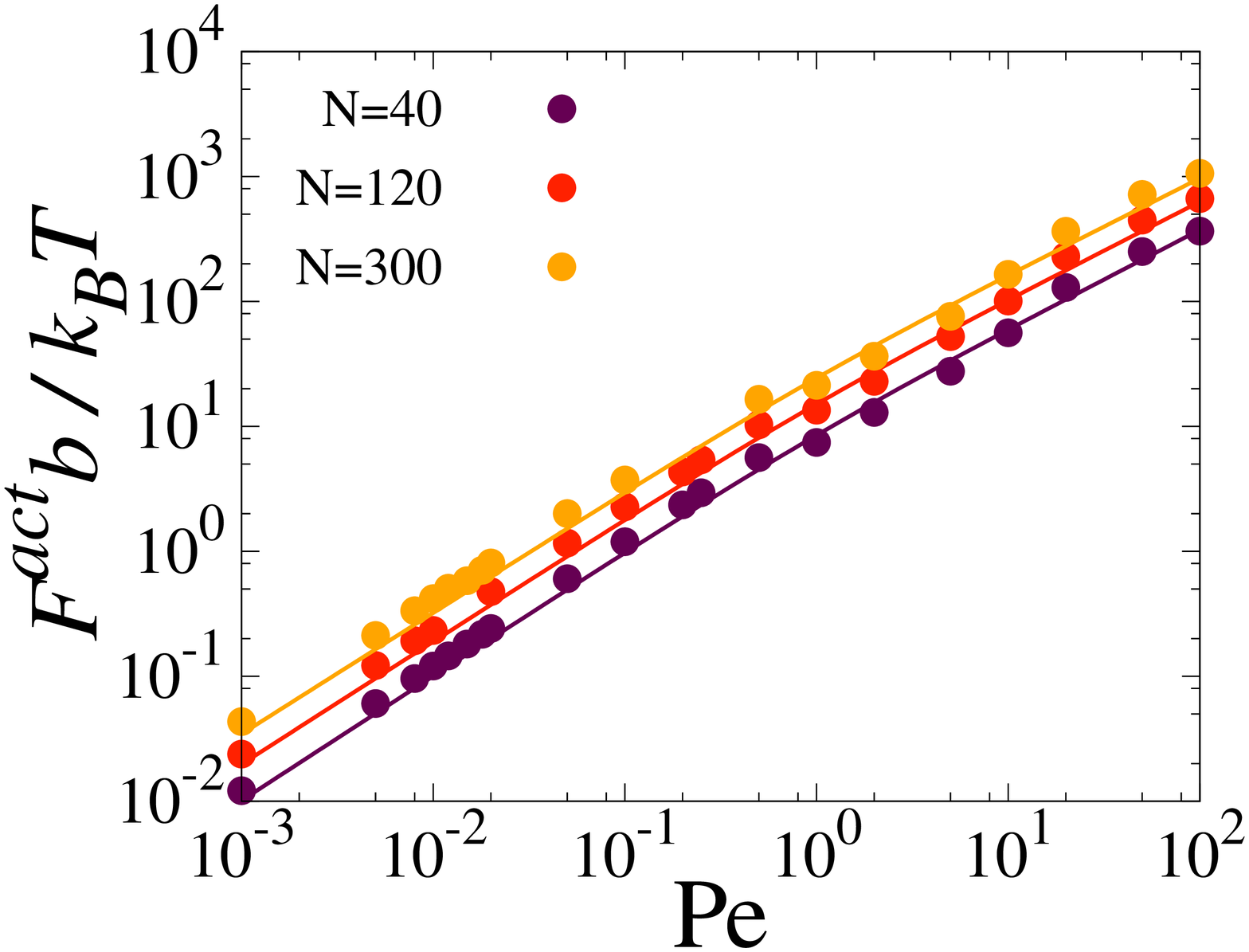}
 \caption{(a) Scaling exponent of of the end-to-end vector as function of $\rm Pe$. The numerical values are consistent, within the statistical error, with the scaling exponent of $R_G$ for $\rm Pe \leq 10$, captured by Eq.(\ref{eq:def-nu}) (solid lines). 
 (b) $R_E$ as function of $\rm Pe$ as obtained from the numerical simulations (points), for $N=30,40,50,60,100,120,160,200,240,300$ (lighter colors standing for larger values of $N$) and their fit (solid lines) with Eq.(\ref{eq:def-R_E}) with $a_{R_E}=1.4$, $c_{R_E}=0.18$ and $h_{R_E}=0.05$. (c) ${ F}^{\rm act}$ as function of $\rm Pe$ as obtained from numerical simulations (points), for $N=40,120,300$ and their fit (solid lines) with Eq.(\ref{eq:def-R_E}) with $a_{R_E}=1.4$, $c_{R_E}=0.18$ and $h_{R_E}=0.05$.}
 \label{fig:scaling_F}
\end{figure}
Eq.(\ref{eq:def-F}) shows that the magnitude of the active force is proportional to that of the end-to-end vector $R_E$. 
The numerical simulations show that $R_E$ retains a power-law dependence on $N$ but with a P\'eclet-dependent exponent, $\nu_{R_E}({\rm Pe})$. Interestingly, for $\text{Pe}<10$ we have $\nu_{R_E}({\rm Pe})\simeq \nu_{R_G}({\rm Pe})$ (see Eq.~(\ref{eq:def-nu}) of the main text), i.e. it is possible to fit both the scaling exponent of $R_G$ and that of $R_E$ with the same function (as shown in Fig. \ref{fig:scaling_F}a.).
For $\text{Pe}\geq 10$, $\nu_{R_E}({\rm Pe})$ shows a non-monotonous dependence on $\text{Pe}$ that we speculate might depend in the fact that in this regime the bead-bead distance increases.
Then, using Eq.(\ref{eq:def-nu}) and the dependence of $R_E=|\mathbf{R}_E|$ on $N$ and $\rm Pe$ extracted from the numerical simulations (Fig.\ref{fig:scaling_F}b) we obtain
\begin{equation}
 R_E=b\dfrac{a_{R_E}+h_{R_E}\ln(\rm Pe)}{\left(\rm{Pe} +1\right)^{c_{R_E}}}N^{\nu(\rm Pe)}
 \label{eq:def-R_E}
\end{equation}
where $a_{R_E}$, $h_{R_E}$ and $c_{R_E}$ are dimensionless coefficients independent of $N$ and $\rm Pe$. By fitting the data we have obtained $a_{R_E}=1.4$, $h_{R_E}=0.05$ and $c_{R_E}=0.18$. We remark that Eq.(\ref{eq:def-R_E}) has the same structure as Eq.(\ref{eq:def-R_G}) in the main text, and we find $a_{R_E} \sim 2a_{R_G}$, $h_{R_E} \sim 2h_{R_G}$ and $c_{R_E}\sim c_{R_G}$. 
We remark that the predictions of Eq.(\ref{eq:def-R_E}), jointly with $\nu_{R_E}({\rm Pe})\simeq \nu_{R_G}({\rm Pe})$, are in quantitative agreements with the numerical data even for $\rm Pe\geq 10$ showing that in this regime $R_E$ is less sensitive on the exact value of the scaling exponent. 
Finally, as a check of our prediction of the dependence of $F^{\rm act}$ on $R_E$, we used Eq.~(\ref{eq:def-R_E}) to fit the data. Interestingly, Fig.\ref{fig:scaling_F}c shows a good agreement between the prediction of Eq.~(\ref{eq:def-F}) and the scaling of $\mathbf{F}^{act}$ extracted from the numerical simulations. 
\paragraph{Dynamics of the center of mass.} 
By summing Eq.~(\ref{eq:eq_brownian}) over all the monomers $i$, we get the equation governing the motion of the center of mass 
 \begin{equation}
  {\bf\dot{r}}_{CM}\equiv \frac{\beta D_0}{N} \sum_i^N {\bf\dot{r}}_i =  \frac{\beta D_0}{N} \left( \boldsymbol{\xi} + \boldsymbol{\eta}  \right) 
 \end{equation}
with 
\begin{equation}
  \boldsymbol{\xi} \equiv {\bf F}^{\rm act}
  \label{eq:def-xi}
\end{equation}
and
\begin{equation}
 \boldsymbol{\eta} \equiv  \sum_i^N \boldsymbol{\eta}_i
\end{equation}
where $\boldsymbol{\eta}$ is the random noise accounting for the thermal fluctuations of the center of mass and $\boldsymbol \xi$, via ${\bf F}^{\rm act} $, accounts for the contributions due to the active forces along the backbone of the polymer. 

The amplitude of the equilibrium fluctuations $\boldsymbol \eta$, is characterized by
\begin{equation}
 \langle  \boldsymbol{\eta} \rangle_\eta=0
\end{equation}
and time correlation 
\begin{equation}
 \langle  \boldsymbol{\eta}(t) \boldsymbol{\eta}(t') \rangle_\eta=2d \frac{(k_BT)^2}{D_0} N\delta(t-t'),
 \label{corr_eta}
\end{equation}
where the factor $N$ arises because $\boldsymbol{\eta}$ is the sum of $N$ independent noises $\boldsymbol{\eta}_i$, each one obeying to the fluctuation-dissipation relation $\langle \boldsymbol{\eta}_i(t)\boldsymbol{\eta}_{i}(t')\rangle = 2dD_0\delta(t-t')$. 
In addition, supported by the presence of the long time diffusive regime, we assume $ \boldsymbol{\xi}$ to be a random force acting on the center of mass with a null average 
\begin{equation}
 \langle \boldsymbol{\xi}  \rangle_\xi=0
\end{equation}
and whose time correlation  depends on time correlation function $C(t)$ of the the end-to-end vector $R_E$ (being the dynamics of $R_E$ related to the slowest polymer mode), i.e. $\langle  \boldsymbol{\xi}(t) \boldsymbol{\xi}(t') \rangle_\xi \propto C(t)$. In the following we approximate $C(t)$ with an exponential function, leading to the following expression
\begin{equation}
 \langle  \boldsymbol{\xi}(t) \boldsymbol{\xi}(t') \rangle_\xi=2d \frac{\xi^2_0}{\tau} e^{-\frac{|t-t'|}{\tau}}
 \label{eq:xi-2}
\end{equation}
where $\xi_0$ has dimensions of a diffusion coefficient and $\tau$ is the correlation time of $R_E$.
Finally, $ \boldsymbol{\eta}$ and $ \boldsymbol{\xi}$ are assumed to be not correlated
\begin{equation}
 \langle  \boldsymbol{\eta}(t) \boldsymbol{\xi}(t') \rangle_{\eta,\xi}=0
\end{equation}

The average displacement after a time $t$ of the center of mass is defined as  
\begin{equation}
\left\langle {\bf r}_{CM}(t)\right\rangle _{\xi,\eta}= \frac{\beta D_0}{N} \left\langle \int_{0}^{t} \boldsymbol{\xi}(t')dt'\right\rangle _{\xi} + \frac{\beta D_0}{N}  \left\langle \int_{0}^{t} \boldsymbol{\eta}(t')dt'\right\rangle _{\eta}=0
\end{equation}
where the last equality is due to the zero-average of $ \boldsymbol{\xi}$ and $ \boldsymbol{\eta}$.

\begin{figure}
\vspace{0.3 cm}
 \includegraphics[scale=0.35]{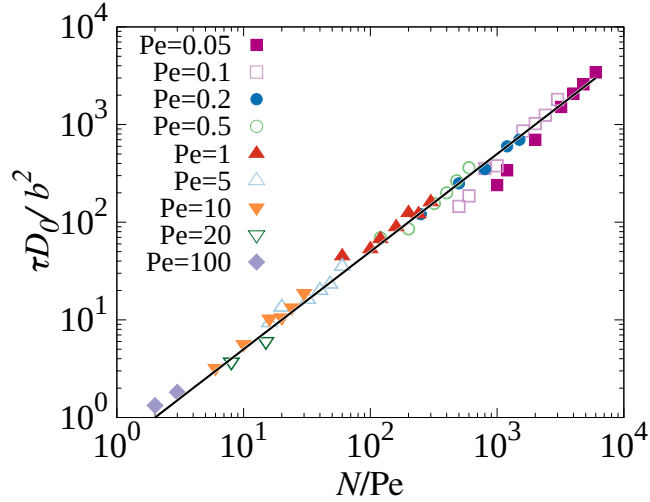}
 \caption{Scaling of the correlation times $\tau$ as function of $N/{\rm Pe}$. The line is a fitting function $\tau = \frac{1}{2} N/\rm Pe$. }
 \label{fig:corr_times}
\end{figure}

Concerning the mean square displacement we have
\begin{eqnarray}
\left\langle {\bf r}^{2}_{CM}(t)\right\rangle _{\xi,\eta} = \dfrac{(\beta D_0)^2}{N^2} \left\langle \int_{0}^{t}\int_{0}^{t}\boldsymbol{\xi}(t')\boldsymbol{\xi}(t'')dt'dt''\right\rangle _{\xi}   + \dfrac{(\beta D_0)^2}{N^2} \left\langle \int_{0}^{t}\int_{0}^{t}\boldsymbol{\eta}(t')\boldsymbol{\eta}(t'')dt'dt''\right\rangle _{\eta}
\label{eq:decorr2}
\end{eqnarray}
Substituting Eq.~(\ref{corr_eta}) and Eq.~(\ref{eq:xi-2}) into the Eq.~(\ref{eq:decorr2}) we get 
\begin{eqnarray}
\left\langle \int_{0}^{t}\int_{0}^{t}\boldsymbol{\xi}(t')\boldsymbol{\xi}(t'')dt'dt''\right\rangle _{\xi}  =  4d\frac{\xi_{0}^{2}}{\tau}\int_{0}^{t}dt'\int_{0}^{t'}e^{-\frac{|t'-t''|}{\tau}}dt'' \nonumber \\
  =  4d\xi_{0}^{2}\int_{0}^{t}dt'e^{-\frac{t'}{\tau}}\left(e^{\frac{t'}{\tau}}-1\right)dt''
  =  4d\xi_{0}^{2}\left(t+\tau e^{-\frac{t}{\tau}}-\tau\right) 
 \label{eq:corr-xi-2}
\end{eqnarray}
and
\begin{eqnarray}
\left\langle \int_{0}^{t}\int_{0}^{t}\boldsymbol{\eta}(t')\boldsymbol{\eta}(t'')dt'dt''\right\rangle _{\eta} & = & 2d \frac{(k_BT)^2}{D_0} N\int_{0}^{t}dt'\int_{0}^{t}\delta(t'-t'')dt'' = 2d \frac{(k_BT)^2}{D_0} N t
 \label{eq:corr-eta-2}
\end{eqnarray}
Concerning the amplitudes of $\xi_0^2$, from Eq.~(\ref{eq:def-xi1}) and Eq.~(\ref{eq:xi-2}) we have
\begin{equation}
 \xi^2_0=\frac{1}{2d}\langle ({\bf F}^{\rm act}(t))^2\rangle\tau
 \label{eq:def-xi1}
\end{equation}
where the correlation time $\tau$ of the end-to-end vector represents also the correlation time of the active force ${\bf F}^{\rm act}$. Assuming $C(t)$ has an exponential decay in time, we extract from the numerical data the correlation times $\tau$, shown in Fig.\ref{fig:corr_times}, from which we obtain the following scaling:
\begin{equation}
 \tau = \tau_0 \frac{N}{\rm Pe}.
 \label{eq:def-tau}
\end{equation}
with $\tau_0=1/2$. We found the previous relation to be valid for $\rm Pe \geqslant 0.05$. 
Substituting Eqs.(\ref{eq:def-F}),(\ref{eq:def-R_E}),(\ref{eq:def-tau}) into Eq.(\ref{eq:def-xi}) we obtain: 
\begin{equation}
 (\beta D_0)^2\xi_0^2 = \left[\frac{1}{2d}\frac{D_0 \tau_0}{b^2}\text{Pe}\frac{\left(a_{R_E}+ h_{R_E}\ln(\text{Pe})\right)^2}{\left(\rm{Pe}+1\right)^{2c_{R_E}}}N^{1+2\nu({\rm Pe})}\right]D_0
 \label{eq:def-xi0}
\end{equation}
Finally, Eq.~(\ref{eq:decorr2}) reads
\begin{equation}
\left\langle {\bf r}^{2}_{CM}(t)\right\rangle_{\xi,\eta} = 2 \left[\frac{D_0 \tau_0}{b^2}\text{Pe}\frac{\left(a_{R_E}+ h_{R_E}\ln(\text{Pe})\right)^2}{\left(\rm{Pe}+1\right)^{2c_{R_E}}}N^{2\nu({\rm Pe})-1}\right]D_0\left(t+\tau e^{-\frac{t}{\tau}}-\tau\right)+2d \frac{D_0}{N} t.
\label{eq:MSD}
\end{equation}
From the previous expression we can compute the long time diffusion coefficient $D_\text{long}$ as
\begin{equation}
 D_\text{long}\equiv \lim_{t\rightarrow\infty} \dfrac{\left\langle {\bf r}^{2}_{CM}(t)\right\rangle _{\xi,\eta}}{2d\, t} = D_0\left[\frac{1}{d}\frac{D_0 \tau_0}{b^2}\text{Pe}\frac{\left(a_{R_E}+h_{R_E}\ln(\text{Pe})\right)^2}{\left(\rm{Pe}+1\right)^{2c_{R_E}}}N^{2\nu({\rm Pe})-1}+\frac{1}{N}\right].
 \label{eq:final-D}
\end{equation}
Eq.(\ref{eq:final-D}) shows that for small values of $\rm Pe$ and $N$ the first term in the brackets can be neglected and the diffusion coefficient scales as $\sim N^{-1}$ as for a passive polymer. By increasing $\rm Pe$ and $N$, the first term becomes dominant and, since the term $2\nu({\rm Pe}) -1$ is generally small, $D_\text{long}$ becomes almost independent of the polymer size $N$. 

\subsection*{Supplementary video}

The video shows a comparison of the motion of an active polymer ($\rm Pe=10^2$,  green on the left) and a passive one (blue o the right). Both polymers have the same size $N=160$ and the terminal monomers are colored in red. Since the motion of the active polymer is faster than the passive one, for sake of visualization the conformations are 
displayed each 50 and 5000 Brownian dynamics time steps for the active and passive polymers, respectively. The video clearly shows that the active polymer tends to assume globule-like (bundle) conformations, characterized by a reduced gyration radius with respect to the passive polymer.

\end{document}